\documentclass{PoS}

\title{Classical Nova Persei 2018 outburst from the dwarf nova V392 Per}

\ShortTitle{Classical nova outburst from the dwarf nova V392 Per}

\author{\speaker{Drahom\'ir Chochol}$^1$, {Sergey Shugarov}$^{1,2}$, {\v{L}ubom\'ir Hamb\'alek}$^1$, {Augustin Skopal}$^1$, {\v{S}tefan Parimucha}$^3$ and {Pavol Dubovsk\'y} $^4$ \\
                 \llap{$^1$}Astronomical Institute, Slovak Academy of Sciences, 059 60 Tatransk\'a Lomnica, Slovakia\\
                 \llap{$^2$}P.K. Sternberg Astronomical Institute, M.V. Lomonosov Moscow State University, Russia\\
                 \llap{$^3$}Institute of Physics, Faculty of Natural Sciences, \v{S}af\'arik University, Jesenn\'a 5, 040 01 Ko\v{s}ice, Slovakia\\
                 \llap{$^4$}Vihorlat Astronomical Observatory, Mierov\'a 4, 066 01 Humenn\'e, Slovakia\\
                 E-mail: \email{chochol@ta3.sk}, \email{shugarov@ta3.sk}, \email{lhambalek@ta3.sk}, \email{astrskop@ta3.sk}, \email{stefan.parimucha@upjs.sk}, \email{vihorlatobs1@stonline.sk}}

\abstract{On 2018, April 29, a bright classical nova (CN) Per 2018 was discovered. Its progenitor is well-known dwarf nova V392
    Per. In this contribution we analyse $UBVR_{C}I_{C}$ photometry and optical spectroscopy of the CN V392 Per. From the $V$ light curve (LC) we found the brightness decline times t$_{2,V}$ = 3 d, t$_{3,V}$ = 10 d and calculated absolute magnitude of the nova at maximum $MV_{max}$ = -9.30 ${\pm}$0.57  using the new $MV_{max}$ - t$_{3}$ "universal" decline law and $MV_{15}$ relations, adopting the Gaia data for CNe. We determined the colour excess $E_{B-V}$ = 0.90$\pm$0.09 and distance to the nova $d$ = 3.55$\pm$0.6 kpc. The optical spectrum obtained in brightness maximum resembles that of the F2 supergiant. Its bolometric luminosity computed by fitting the continuum by atmospheric and black-body models is in agreement with the luminosity, that we have found from photometry. We estimated the mass of the ONe white dwarf in V392 Per as $M_{wd}$ = 1.21 M$_{\odot}$. The CN Per 2018 can be classified as a fast super-Eddington nova with an outburst LC of plateau type. Nova displayed  He/N spectrum classification, large expansion velocities, and triple-peaked emission-line profiles during the decline, explained by equatorial ring seen nearly face on and a bipolar flow aligned almost with the line of sight. The post maximum spectra of CN Per 2018 and available radio data were used to estimate the inclination angle of the system as $i\sim$ 9$^{\circ}$. The difference in intensity of redward and blueward emission bumps is possible to explain by about 1.5 times higher density of the receding outtflow. The rapid increase of the bipolar outflow radial velocities by $\sim$300 km/s around day 5 after the maximum was caused by the fast bipolar winds from the burning white dwarf after shrinking of its pseudophotosphere.}

\FullConference{The Golden Age of Cataclysmic Variables and Related Objects V (GOLDEN2019)\\
		2-7 September 2019\\
		Palermo, Italy}

\begin{document}

\section{Introduction}

CNe are cataclysmic variables (CVs) with 6 to 19 mag brightness increase, caused by thermonuclear runaway (TNR) event on the surface of a
    white dwarf (WD). They arise in close binaries with orbital periods of a few hours up to $\sim$ 6 days, consisting of a red dwarf or cool subgiant filling up its Roche-lobe and a mass-accreting WD.  While the luminosity of the binary is close to 1\,L$_\odot$, during the outburst can reach a few times 10$^5\,L_\odot$. After the TNR, the photosphere of the WD component of the CN expands to supergiant dimensions and engulfs the binary. Due to a strong wind, a large part of the envelope is ejected and the photospheric radius shrinks.

Classical novae can be divided according to their photometric appearance to the fast and slow. Classification is usually based on a time interval
    in which nova fades by 2 or 3 magnitudes ($t_2$, $t_3$) from its maximum brightness. The fast super-Eddington novae ($t_2 <$ 13, $t_3 <$ 30 days) have smooth LCs with well-defined maxima. The slow Eddington novae ($t_2 >$ 13, $t_3 >$ 30 days) have structured LCs and many of them have standstills at maximum and dust formation at later stages [9]. According to the properties of the LCs  during nova declines, [28] proposed seven types of LCs:  S (smooth), P (plateau), D (dust dip), C (cusp), O (oscillations), F (flat-topped), and J (jitter).

The spectra display either He+N or Fe II emission lines as the most prominent non-Balmer lines at maximum light. Fe II spectra are formed in a large
    circumbinary envelope of gas, whose origin is the secondary star, while He/N spectra are formed in WD ejecta. In hybrid objects, both classes of spectra appear sequentially due to changing parameters in the two emitting regions [32].

Dwarf novae (DNe) are CVs  with 1-6 mag brightness increases and durations of days-to-weeks. Osaki [23] proposed a unification model
    for DN outbursts within the framework of the disk instability model, in which two different intrinsic instabilities (the thermal and the tidal
    instability) within accretion disks play an essential role.

The donor in CNe, recurrent novae (RNe) and DNe can be either a red dwarf (for orbital periods of the order of hours) or a cool subgiant (for orbital
    periods in the range 10 hrs to 6 days).

\section{Discovery of the CN Per 2018 and its progenitor DN V392 Per}

Classical Nova Persei 2018 was discovered as a new transient TCP J04432130+4721280 by Nakamura [22] on
    2018 April 29.474 UT at mag 6.2. He noted that the transient is spatially coincident with U Gem type dwarf nova (DN) V392 Persei [10]. According to our data,
    the nova reached the brightness maximum  $V_{max}$ = 6.24 mag, $B_{max}$ = 7.12 mag on 2018 April 29.83 (JD  2458238.33), taken as a zero point (0.0 d$_{AM}$) for counting the times of photometric and spectroscopic observations through our paper.

Darnley \& Starfield [8] discussed the available data of Nova progenitor V392 Per. The 2004-2018 AAVSO LC
    indicates the quiescent system with V $\sim$ 16-17 mag and a few dwarf nova outbursts with the amplitude  2 - 3.5 mag, the last one in 2016. Gaia Data Release 2 (DR2; [11],[12]) contains a parallax measurement for V392 Per of 0.442${\pm}$0.053 mas, which could indicate a distance of 3.9$^{+1.0}_{-0.6}$ kpc. The 3D dust maps of Green et al. [13][14] yield a reddening of $E_{B-V}$ = 0.9${\pm}$0.1 over the Gaia distance range. Schaefer [24] and Bailer-Jones et al. [1] used Gaia DR2 data to find the distance of V392 Per to be  4161$^{+2345}_{-440}$ pc and 3416$^{+750}_{-533}$ pc, respectively. Darnley and Starfield [8] found that the SED of V392 Per in quiescence is similar to the CN GK Per or RNe U Sco and M31N 2008-12a, that contain evolved donors (subgiant or red-clump giant).

\section{ATels spectroscopy and multifrequency observations of the CN Per 2018}

Classical nova outburst was confirmed spectroscopically [31] on April 30.116 UT (0.286 d$_{AM}$) with
    the 2.4-m Hiltner telescope on Kitt Peak MDM Observatory. The spectrum exhibits broad H$\alpha$ emission, with FWHM 5200 km~s$^{-1}$ and absorption displaced by - 2680 km~s$^{-1}$ with respect to the fitted center of emission, and Fe II P Cygni-type line profiles. The spectrum is similar to the spectrum taken by R.\,Leadbeater (ARAS) at April 29.898 UT (0.068 d$_{AM}$).

Li et al. [15] detected nova as a strong gamma-ray source using the data collected on April 30 (00:00:00 - 23:25:53) by Fermi Large
    Area Telescope, covering the energy range from 20 MeV to more than 300 GeV.

The high-resolution spectra (R $\sim$ 30 000) of the nova, obtained by Tomov et al. [29] on May 1.78 (1.95 d$_{AM}$) and May 2.78
    (2.95 d$_{AM}$) with the 2-m telescope at Rozhen observatory exhibited broad emission lines with complex shape. The strongest emissions are H$\alpha$, H$\beta$, Fe II 42 multiplet, Ca II IR triplet and O I lines. The FWHM for H$\alpha$, H$\beta$ emissions are 5600$\pm$200 km/s. Three emission peaks at about -2000 km/s, -250 km/s and 1900 km/s, and weak P Cyg absorptions at -4000 km/s in H$\alpha$, while -3700 and -4100 km/s in H$\beta$ are detected. The extinction derived from 5780 \AA, 5797 \AA, 6614 \AA~ diffuse interstellar bands and KI line is $E_{B-V}$ = 1.18$\pm$0.1 mag.

The structured Balmer line profiles with three individual peaks were detected also in the low-resolution spectra (R $\sim$ 2200 and R $\sim$ 2600) taken by the 2-m Liverpool telescope (LT) on May 2.86 UT (3.03 d$_{AM}$) [6].

Linford et al. [16] reported radio observations with Karl G. Jansky Very Large Array (VLA) in Soccoro and  Arcminute Microkelvin Imager Large
    Array (AMI-LA) in Cambridge. The nova was clearly detected with VLA at 5.0 and 7.0 GHz on 2018, May 8.7 (8.87 d$_{AM}$) and May 12.6 (12.77 d$_{AM}$)  and with AMI-LA at 15.5 GHz on 2018, May 11.5 (11.67 d$_{AM}$) and May 13.6 (13.77). The nova is increasing in flux density at all frequencies. A brightness temperature of the order of 10$^5$ K, estimated from radio data, the ejecta expansion velocity of 4000 km/s [29] and a distance of 3.9 kpc [8] is higher, than expected for nova ejecta emitting via thermal bremsstrahlung. Therefore, they suspected a significant contribution from synchrotron emission, related to the shocks that produced the population of accelerated particles necessary for the detected gamma-ray emission [15]. During 2018, May 17 - 21 compact knots (with brightness temperature in excess of  10$^7$ K)  moved away from each other with a projected  velocity of 1350 km/s [17].

The LT spectra of the nova taken after the sun constrain on 2018, July 13.21 UT (74.38 d$_{AM}$) showed that the object entered the post-nova
    eruption nebular phase [4],[5]. The spectra exhibited H$\alpha$, H$\beta$, H$\gamma$, H$\delta$, He I and He II (4686 \AA) triple-peaked emission lines and extremely strong double-peaked [O III] lines at 4363, 4959 and 5007 \AA.  The width of the [O III] lines are consistent with those of H I lines. The central narrow ($\sim$ 50 km/s) and bright emission peak of H$\alpha$ line is accompanied by fainter and broad blueward and redward emission bumps with FWHM of $\sim$ 1600 km/s, shifted by $\sim \pm$ 1750 km/s. The blueward bump is ~2/3 the height of the redward bump.

Darnley et al. [7] detected the nova as an X-ray source on 2018, July 20 (81 d$_{AM}$), when it could be pointed by the Swift Observatory after the sun
    constrains. The Swift UVOT photometry in v,b,u,w1,m2,w2 filters was obtained on 2018, July 20, 26, 27. The large flux deficit  through the m2 filter suggests a high extinction toward the system.

The spectra with the resolution R $\sim$ 1100, obtained by Munari \& Ochner [19] on Aug. 9.061 UT (101.23 d$_{AM}$ ) using
    the Asiago 1.22-m telescope, show the transition to advanced nebular conditions  with [OIII] 4959+5007  flux 3.6 times the integrated flux of H$\alpha$ + [NII]. All emission lines were composed of two broad and well-separated components (both with a FWHM ~1800 km/s). The bluer component was generally the strongest. The velocity separation of these broad components ranges from 3400 km/s for H$\alpha$ up to 4500 km/s for [NeV] 3426~ \AA. In most of the emission lines, except [NeV], the central narrow and low-velocity component is detected. The peak intensity of this central, narrow component (Nw) varies greatly from line to line compared to the intensity of the broad components (Bc), ranging from Nw $>>$ Bc for HeII 4686, to Nw $>$ Bc for HeI lines, Nw $\sim$ Bc for Balmer lines and Nw $<<$ Bc for [NeIII], [OIII], [NII] nebular lines. Such profiles for the emission lines are closely similar to those observed at late stages in the extremely fast and He/N V2672 Nova Oph 2009, modelled by Munari et al. [20] as a bipolar flow aligned with the line of sight plus an equatorial torus seen face-on. The large intensity of the [NeV] 3426 line, comparable in intensity to H$\alpha$ + [NII], suggests that V392 Per is a Neon Nova.

The outburst of CN Per 2018 has begun to wane during the sun constraint between April 26, 2019 and July 18, 2019, when optical spectra dominated by broad nebular forbidden lines of [O III] at 4959 and 5007 decreased in intensity [21].

\section{Our photometry and basic parameters of the CN Per 2018}
Our $U,B,V,R_{C},I_{C}$ CCD observations of the nova were obtained at the Star\'a Lesn\'a Observatory of the Astronomical Institute of the Slovak
    Academy of Sciences (AISAS) (Zeiss 60-cm telescope, Maksutov 18-cm telescope, 6-cm telephoto lens) and the Kolonica Observatory (100-cm Vihorlat National Telescope (VNT) and 3 other small telescopes of the Vihorlat Astronomical Observatory, Humenn\'e, Schmidt-Cassegrain 50.8-cm telescope of the \v{S}af\'arik University, Ko\v{s}ice). The data were processed by a standard way using the nearby comparison stars and reduced to the standard Johnson-Cousins system. Our data were complemented with those available from AAVSO, VSNET and other sources. The $B$ and $V$ data around the brightness maximum and during 16 days after maximum are presented in Fig. 1. The nova displays a photometric plateau phase between day 5 and 8 after maximum, so the LC is of a type P (Plateau) according to classification [28]. The available $U,B,V,R_{C},I_{C}$ data and corresponding colour indices during 38 days after maximum are presented in Fig. 5.

\begin{figure}
\centerline{\includegraphics[width=.95\textwidth]{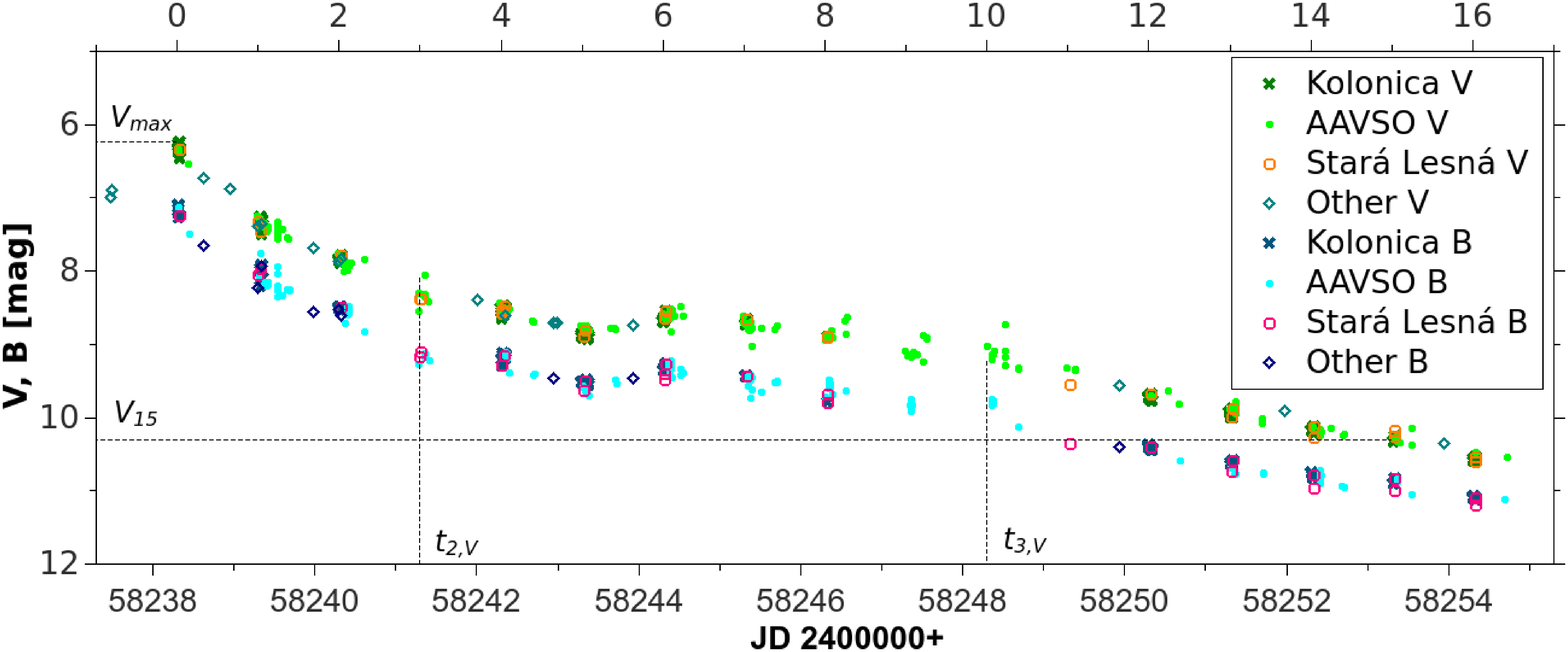}}
\caption{$V$ and $B$ LCs of the CN Per 2018 in the first 16 days after maximum brightness}
\label{fig1}
\end{figure}

    From the $V$ LC we found the brightness decline after maximum $V_{max}$ = 6.24 mag characterized by the times $t_{2,V}$ = 3 d, $t_{3,V}$ = 10 d, and the $V$ magnitude at day 15, $V_{15}$ = 10.3 mag ($V_{max}$ - $V_{15}$ = - 4.06 mag). Selvelli and Gilmozzi [25] published new "universal" decline law relation $MV{_{max}} = (-11.08\pm0.33) + (2.12\pm0.20)\log{t_{3}}$ and $MV_{15} = -5.58{\pm0.41}$ empirical relation, based on which we estimated the absolute magnitude of the CN Per 2018 in maximum as $MV_{1,max}$ = -8.96 ${\pm}$0.53 and $MV_{2,max}$ = -9.64 ${\pm}$0.41, respectively. The unweighted average is $MV_{max}$ = -9.30 ${\pm}$0.57 mag. The relation in  [24]: $M_{wd} = 1.384 - 0.367log{t_{2}}$  allows to find the mass of the WD in the binary as $M_{wd}$ = 1.21 M$_{\odot}$.

    The interstellar extinction towards the CN Per 2018 can be found using our photometric $B,V$ data and methods described in [30]: i) from the comparison of the observed CN Per 2018 colour index $(B-V)_{max}$ = 0.88, affected by extinction, with the intrinsic colour index for novae $(B-V)_0$ = 0.23 $\pm$ 0.06, we have $E_{B-V}$ = 0.65 $\pm$ 0.06, ii) all novae 2 mag below maximum have an intrinsic colour index  $B-V = - 0.02{\pm0.04}$, so the value $B-V$ = 0.83 found for CN Per 2018 yields $E_{B-V}$ = 0.85,
    iii) Tomov et al. [29] determined the reddening $E_{B-V}$ = 1.18${\pm}$0.10 from the spectra taken a few days after the outburst from the interstellar K I (7698.979 \AA~) line and diffuse interstellar bands at 5780 \AA, 5797 \AA, and 6614 \AA,
    iv) we have used the 3D dust map [13] and the distance of V392 Per 3416$^{+750}_{-533}$ found by [1] to
    determine  $E_{B-V}$ = 0.91$^{+0.4}_{-0.5}$. The mean value of the reddening found from the data mentioned above
    is $E_{B-V}$ = 0.90${\pm}$0.09. Corresponding absorption is Av = 2.79${\pm}$0.28. This value and the distance modulus of the nova $V_{max}-MV_{max}$ = 15.54${\pm}$0.2 yields a corresponding distance to the nova of 3.55${\pm}$0.6 kpc.

Schaefer [24] mentioned that ``The many variations on the  `maximum-magnitude-rate-of-decline' (MMRD) relation are all found to be poor, too poor to be useable, and even to be non-applicable for 5-out-of-7 samples of nova, so the MMRD should no longer be used.'' According to the measure of agreement between the distances found from Gaia and from MMRD relations, he classified the novae as gold, silver and bronze sample. His data for silver case V392 Per were as follows: $V_{max}$ = 5.6 mag, $t_{2}$ = 2\,d, $t_{3}$ = 4\,d, $V_{15}$ = 10.2 mag, $A_{V}$ = 1.6 mag, $M_{max}$ = - 9.0 mag. The distance determined from these data is 3981 pc, close to the distance 4161$^{+2345}_{-440}$ pc found by Schaefer [24] from Gaia data. The source of discrepancy between the Schaefer's  and our results is the fact, that he used the value of brightness maximum 5.6 mag from an unfiltered image taken by D.\,Buczynski [2] at  JD 2458238.4 (see also https://britastro.org/node/13057). As seen from our data, the $V$ magnitude of the nova in maximum was only 6.24 mag.

\begin{figure}
\centerline{\includegraphics[width=.60\textwidth,angle=270]{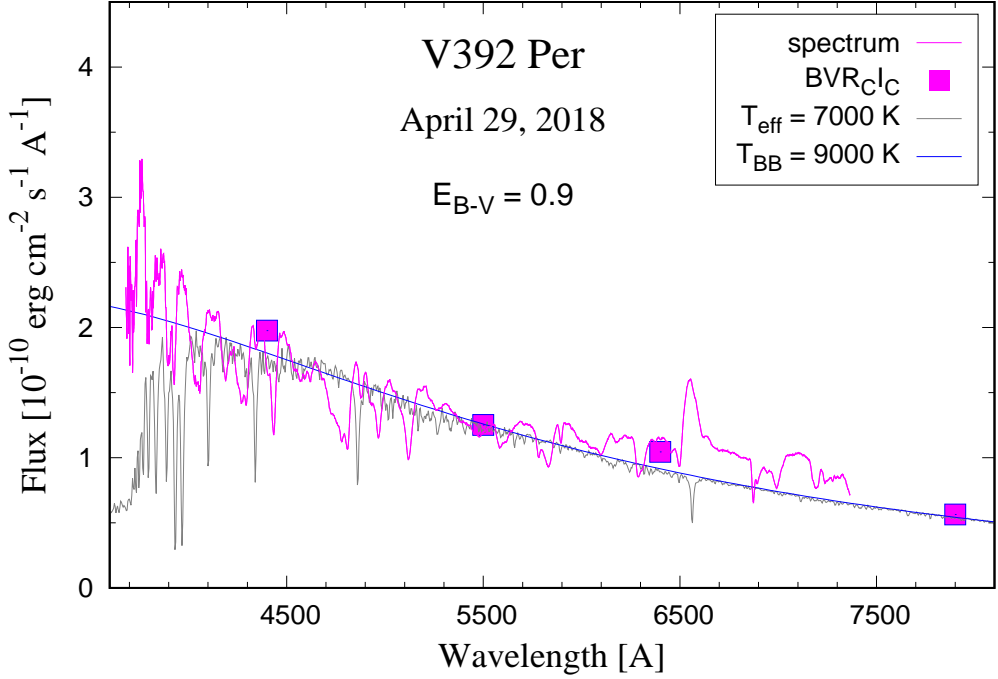}}
\caption{ The spectrum of V392~Per (in magenta) taken at the optical maximum of the nova on April 29.89, 2018, scaled to $BVR_{\rm C}I_{\rm C}$ fluxes (squares). The continuum can be matched by the atmospheric model (gray) and/or by the blackbody radiation (blue).
Observations are dereddened with $E_{B-V} = 0.9$\,mag.}
\label{fig2}
\end{figure}

Basic parameters of CN Per 2018, we derived from photometric observations, are independently supported by modelling the optical spectrum obtained at the maximum brightness of the nova. Fig. 2 shows the low-resolution spectrum (R$\sim$530, 368--737\,nm) taken by R.\,Leadbeater (ARAS) on April 29.894 UT (0.064d$_{AM}$) near the optical maximum. The spectrum is dereddened with $E_{B-V} = 0.9$\,mag and scaled to simultaneous $BVR_{\rm C}I_{\rm C}$ photometric fluxes. The continuum spectrum resembles that produced by a star of spectral type F2. Using the grid of atmospheric models [18] we matched the continuum by a synthetic spectrum characterized with the effective temperature $T_{\rm eff} = 7000$\,K and the scaling factor $\theta_{\rm WD} = R_{\rm WD}^{\rm eff}/d \sim 2.36\times 10^{-9}$, which is the angular radius of the WD pseudophotosphere given by its effective radius $R_{\rm WD}^{\rm eff}$ (= the radius of a sphere which has the same luminosity as the observed stellar photosphere) and the distance $d$. These fitting parameters correspond to $R_{\rm WD}^{\rm eff}\sim 356\,(d/3.4\,{\rm kpc})$\,R$_{\odot}$
    and the luminosity of the WD $L_{\rm WD}\sim 1.1\times 10^{39}\,(d/3.4\,{\rm kpc})^2$\,erg\,s$^{-1}$. The continuum spectrum is also
    well comparable to that of a blackbody radiating at the temperature $T_{\rm BB}\sim 9000$\,K, scaled with
    $\theta_{\rm WD}\sim 1.7\times 10^{-9}$, which correspond to $R_{\rm WD}^{\rm eff}\sim 258\,(d/3.4\,{\rm kpc})$\,R$_{\odot}$
    and $L_{\rm WD}\sim 1.5\times 10^{39}\,(d/3.4\,{\rm kpc})^2$\,erg\,s$^{-1}$. When we accept the distance 3.55 kpc, the bolometric magnitudes of the outbursted WD for the atmospheric and blackbody models are  $M^a_{bol}$ = -9.0 mag and  $M^b_{bol}$ = -9.33 mag, respectively. Their absolute visual magnitudes $MV^a$ = -8.85 mag and $MV^b$ = -9.18 mag, are close to the values, which we derived from photometry.

\section{Our spectroscopy of the CN Per 2018 and its expanding shell}

Our optical echelle spectra of the CN Per 2018 were obtained on 2018, May 2-6 (3 to 7 days after maximum) with the 60-cm telescope in the G1 pavilion of the AISAS observatory at Star\'{a} Lesn\'{a} (three spectra with the resolution  $R$ $\sim$ 12 000) and 1.3-m telescope of the AISAS observatory at Skalnat\'{e} Pleso (three spectra with the resolution $R$ $\sim$ 24 000). Our data were completed by the Astronomical Ring for Access to Spectroscopy (ARAS) data obtained by small telescopes and spectrographs (Alpy 600, LISA, LHIRES, eShel) covering the region 3800-7200 \AA~ and S/N 50-100. They are available at http://www.astrosurf.com/aras/Aras$\_$DataBase/Novae/2018$\_$NovaPer2018.htm. In the first 38 days after maximum, we have used 42 spectra with the resolution $R\sim$ 500 - 800 and 8 spectra with the resolution $R \sim$ 5000 - 20000 covering the H$\alpha$ region.

\begin{figure}
\centerline{\includegraphics[width=.9\textwidth]{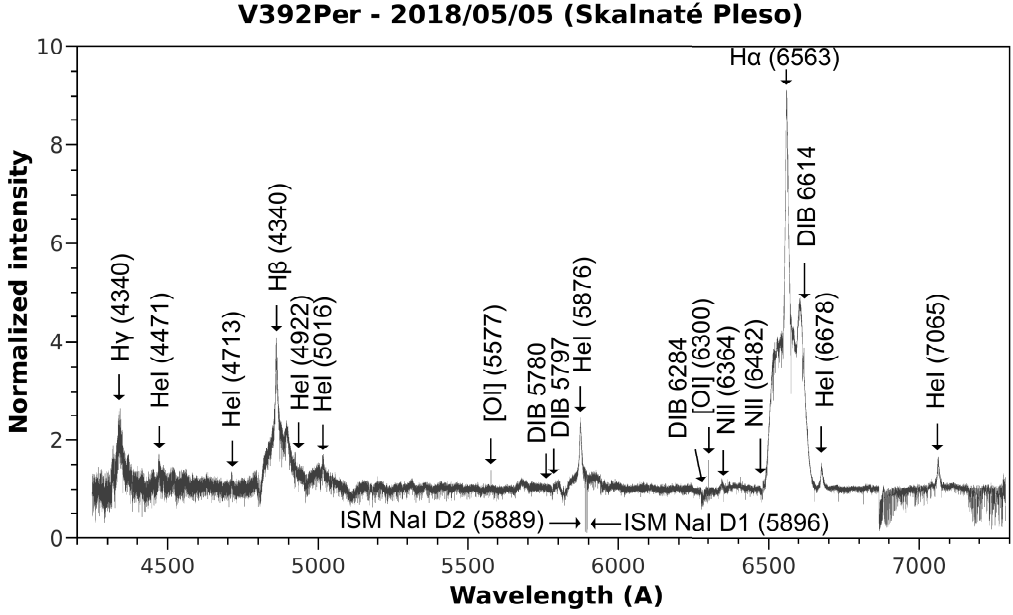}}
\caption{The spectrum of the CN Persei 2018 with identified features. Continuum is normalized to the intensity 1}
\label{fig3}
\end{figure}

The spectrum taken by our 1.3-m telescope 6 days after the maximum is presented in Fig. 3. It clearly shows the structured Balmer H$\alpha$ -
     H$\gamma$ and He I (5876 \AA~) line profiles with three individual peaks, HeI, [OI] and NII lines and strong interstellar NaI lines and DIBs. The sequence of the H$\alpha$ and H$\beta$ line profiles of the CN Per from day 3 to day 7 after the brightness maximum in the radial-velocity scale is presented in Fig. 4. It is clearly seen, that the intensity of the redward broad emission bump is higher than the intensity of the blueward one.

\begin{figure}
\centerline{\includegraphics[width=.90\textwidth]{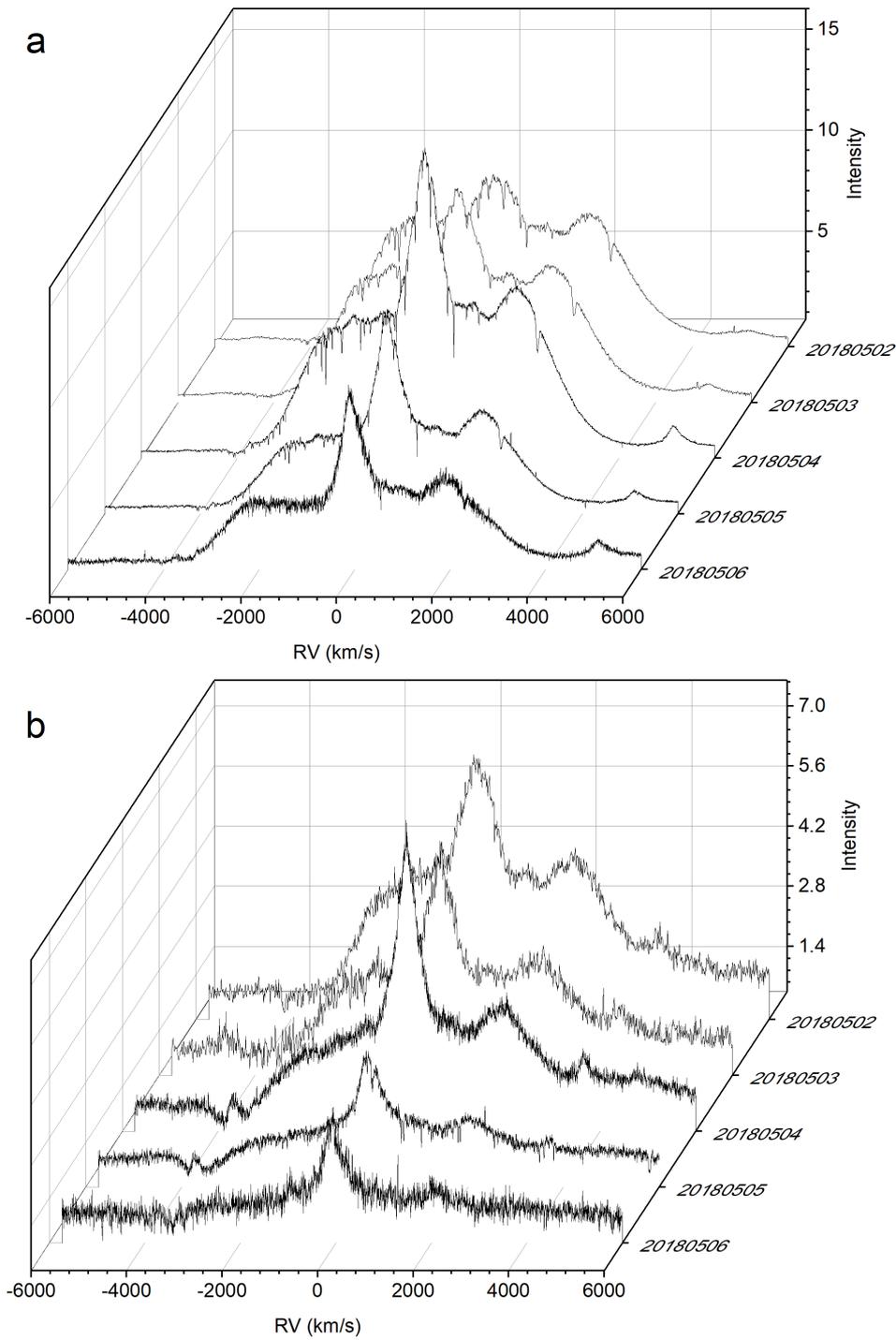}}
\caption{The sequence of the CN Per 2018 H$\alpha$ (a) and H$\beta$ (b) line profiles on high dispersion spectra obtained at Skalnat\'e Pleso and
   Star\'a Lesn\'a Observatory}
\label{fig4}
\end{figure}

For fitting of H$\alpha$ line profile and continuum, we used the code \texttt{fityk} [33]. It is a general-purpose
       peak fitting software with a graphical interface. It offers many non-linear functions, subtraction of continuum, and can fit an arbitrary number of peaks simultaneously by the Levenberg-Marquardt least square method. We selected the region of spectrum around the H$\alpha$ line and fitted continuum by the second-order polynomial. We added a few Gaussian peaks based on the shape of the line, emission, and absorption features. The main emission peak was selected to cover the shape of the wings and the intensity at the central wavelength. We have calculated 3$\sigma$ distance from $\lambda_C$ and computed corresponding radial velocity to $\lambda_C$-3$\sigma$ while using the laboratory wavelength for H$\alpha$. Then the full width at zero intensity (FWZI) of the H$\alpha$ profile,  half-width at half maximum (HWHM) of the central emission and differences in RVs between the central emission and  emission bumps were found and expressed in the radial-velocity scale. They are presented in Fig. 5.

An increase of the $B,V$ magnitudes at 5-6 d$_{AM}$ was accompanied by the increase of FWZI of the H$\alpha$ profile with a maximum at 7-8 d$_{AM}$.
      Due to the fact, that the HWHM of the central emission peak during this period continuously decreases, the increase of FWZI was caused by broadening of emission bumps of the H$\alpha$ profile. The absolute values of radial velocity differences between the central peak and emission bump rapidly increased for about 300 km/s at 5 d$_{AM}$ and lasted till 12 d$_{AM}$. At the same time, we detected the decrease in intensity of the H$\alpha$ emission.

\begin{figure}
\centerline{\includegraphics[width=.9\textwidth]{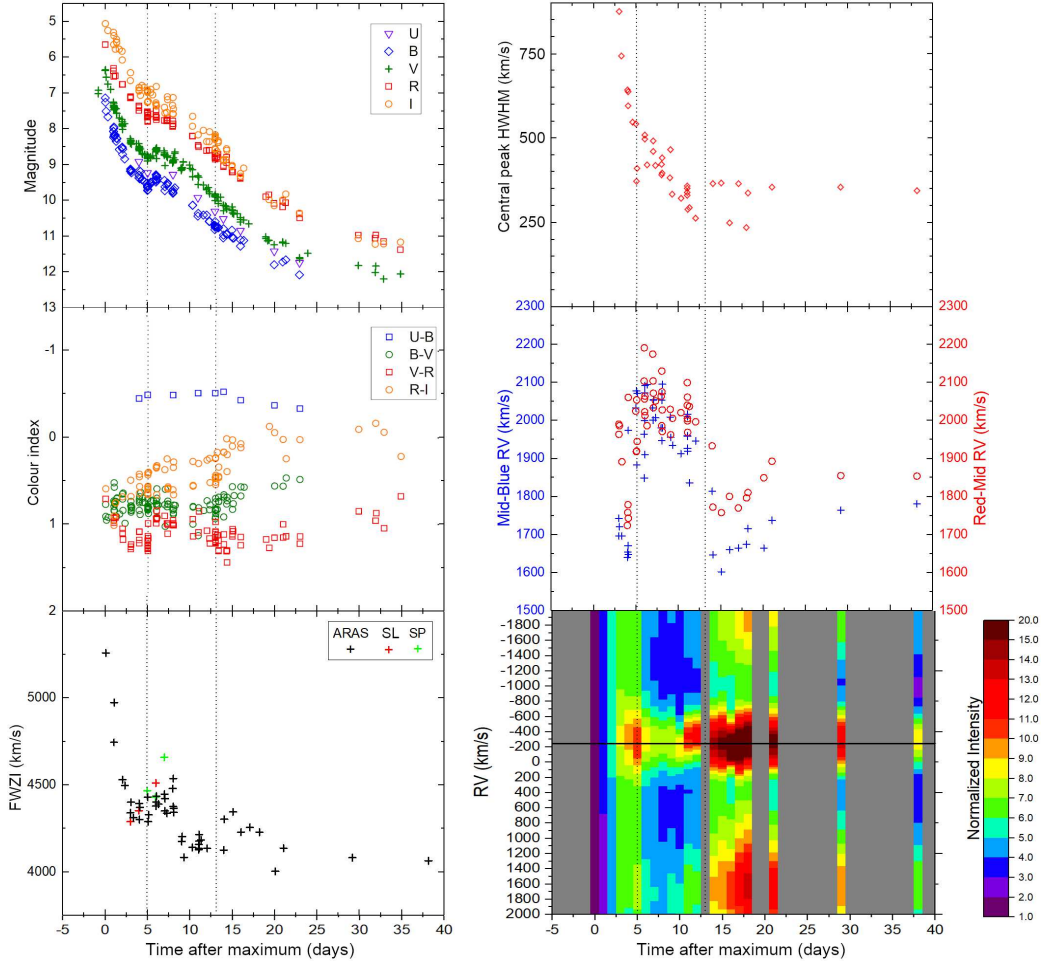}}
\caption{The comparison of the $U,B,V,R_{C},I_{C}$ photometry, colour indices, FWZI of the H$\alpha$ profile, HWHM of the central emission peak, differences in RVs between the central emission peak and emission bumps and intensities of the H$\alpha$ profile of the CN Per 2018 during the first 38 days after maximum}
\label{fig5}
\end{figure}

Chochol et al. [3] derived a kinematic model of an expanding shell of the fast CN V1974 Cygni using the optical and UV spectra, HST and radio
   images. The shell consists of two major components: an outer fast tenuous low-mass envelope and an inner slow main high-mass envelope, accelerated  by slower spherical and faster polar winds. The outer envelope (detected spectroscopically and on radio images) is spherical except for the polar region, where an outflow with larger velocity is observed. The most pronounced features of the inner envelope (detected spectroscopically and on HST images) are an expanding equatorial ring and polar blobs.

\begin{figure}
\centerline{\includegraphics[width=.8\textwidth]{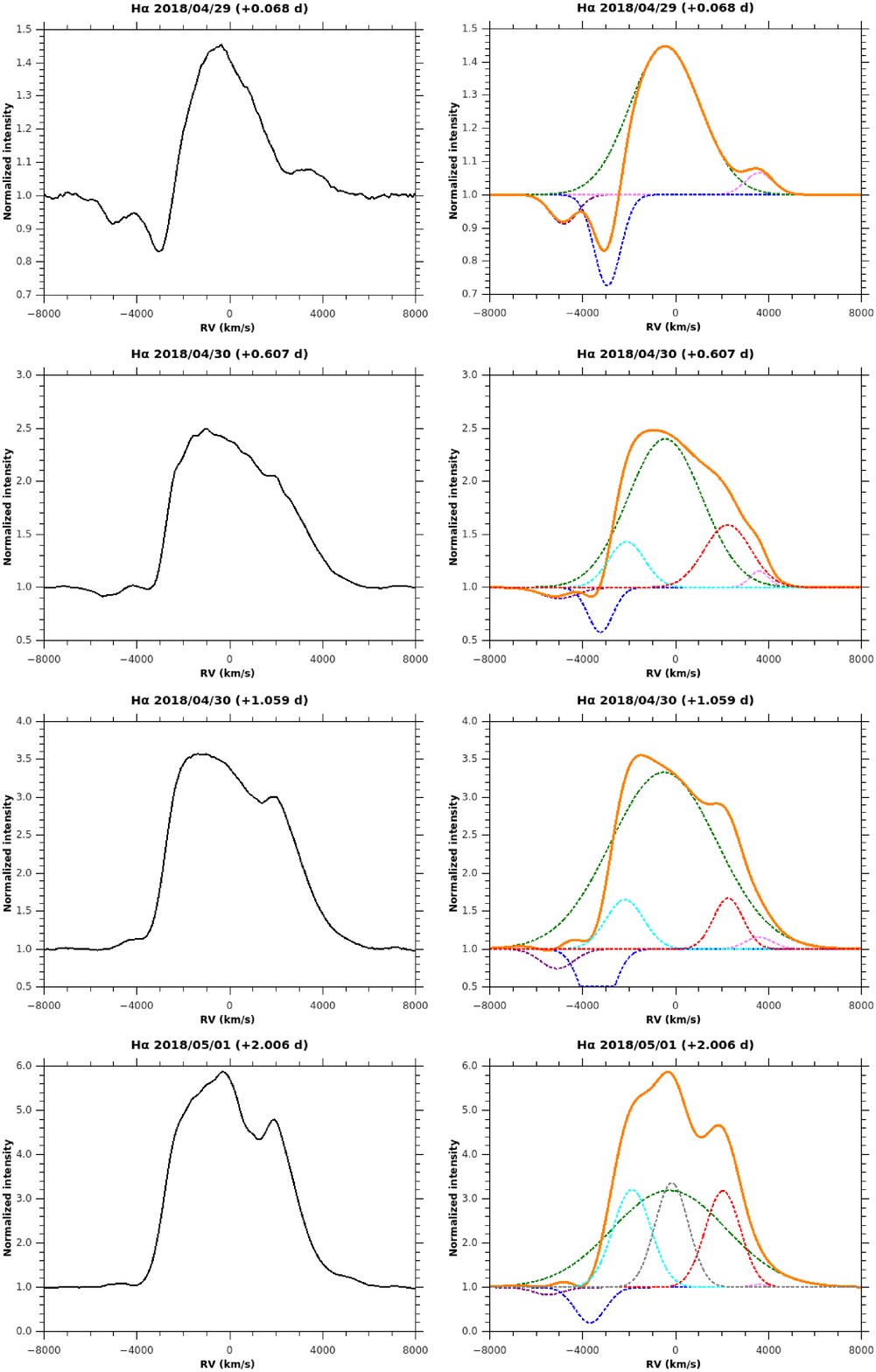}}
\caption{Disentangling of the H$\alpha$ profiles of the CN Persei 2018 in the first two days after maximum. The total profile (TP), ellipsoidal nebula (EN) and components of the outer envelope (OE) and inner envelope (IE) are distinguished by different colours: TP (orange), EN (green), approaching  spherical OE (blue), approaching polar outflow OE (magenta), receding polar outflow OE (pink), equatorial ring IE (grey), approaching polar outflow IE (cyan), receding polar outflow IE(red)}
\label{fig6}
\end{figure}

The similar structure of the shell of CN V392 Per is possible to reveal by disentangling of the H$\alpha$ line profile in the early stages of the
   outburst using the low dispersion spectra presented in Fig. 6. They were obtained by members of the ARAS group R. Leadbeater, Three hills Observatory, UK (0.068 d$_{AM}$, 1.059 d$_{AM}$) and E. Bertrand, St. Sordelin, France (2.006 d$_{AM}$). The spectrum at 0.607 d$_{AM}$ was obtained by M. Fujii (Fujii Kurosaki Observatory, Japan). We detected the outer envelope, shaped by spherical wind during the brightness maximum (day 0.068) as an H$\alpha$ P-Cyg absorption at $\sim$ 2600 km/s (blue line) from the central emission peak located around - 400 km/s (green line). The approaching and receding polar outflow is visible as an H$\alpha$ P-Cyg absorption and an emission bump. Their radial velocities show about $\pm$ 4200 km/s displacement from the central emission peak. Both components were detected by [17] also on radio maps of the nova from May 17 and 21, 2018 (18 and 22 d$_{AM}$) as two compact knots with brightness temperatures in excess of 10$^7$ K, which move away from each other with a projected velocity of 1350 km/s using a distance of 3.9 kpc [8]. The radial velocity of the outflow 4200 km/s and its tangential velocity 675 km/s allows estimating the inclination angle of the object as $i\sim$ 9$^{\circ}$.

\begin{figure}
\centerline{\includegraphics[width=.3\textwidth]{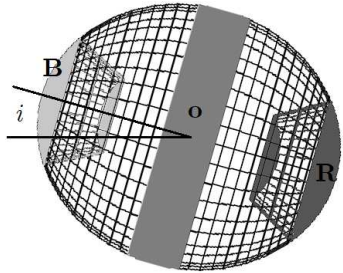}}
\caption{The SHAPE model with equatorial ring (o) and polar blobs (B,R)}
\label{fig7}
\end{figure}

The structure of the ejecta consisted of equatorial ring and polar blobs was find by Munari et al. [20] in the extremely fast nova V2672 Oph (Nova Oph 2009). Their morpho-kinematical modelling of the triple-peaked H$\alpha$ profile of the nova using the code SHAPE2 [26],[27] showed that the system is seen pole-on, with an inclination of 0$^{\circ}$ $\pm$ 6$^{\circ}$ and an expansion velocity of the polar blobs 4800$^{+900}_{-800}$ km/s. A bipolar flow is aligned with the line of sight and an equatorial torus seen face-on.

\begin{figure}
\centerline{\includegraphics[width=.8\textwidth]{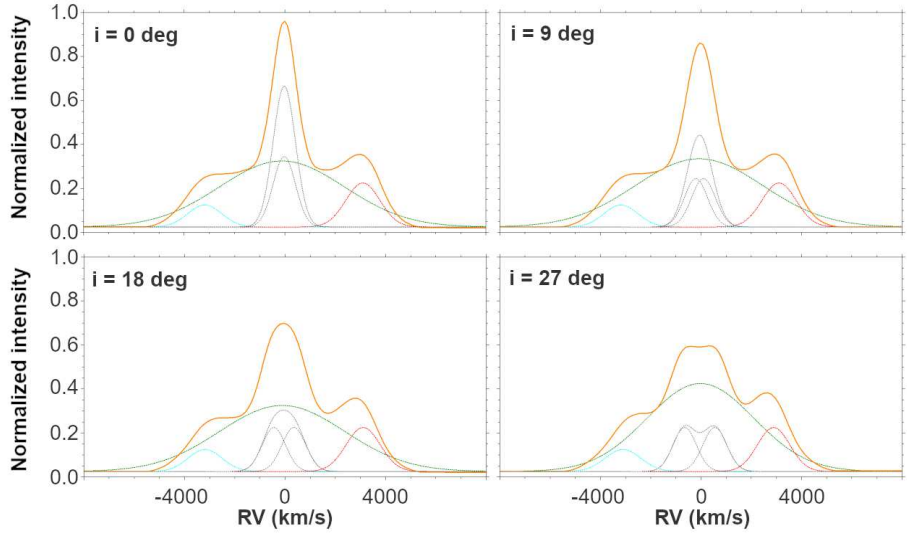}}
\caption{The SHAPE model fits. The total profile (TP), ellipsoidal nebula (EN) and components of the inner envelope (IE) are distinguished by different colours: TP (orange), EN (green), equatorial ring IE (grey), approaching polar outflow IE (cyan), receding polar outflow IE(red)}
\label{fig8}
\end{figure}

We applied the morpho-kinematical code SHAPE2 [26], [27]  for analysis and disentangling the three-dimensional geometry and kinematic structure of the outburst H$\alpha$ spectra of V392 Per. The SHAPE model is presented in Fig. 7. The equatorial ring was detected on triple-peaked H$\alpha$ profiles as a central emission and polar blobs as blueward and redward bumps. The line profiles modelling presented in Fig. 8, confirm a small inclination angle of the object ($i \sim$  9$^{\circ}$). The asymmetric outflow can be caused by $\sim$ 1.5 times higher density of the outflowing material receding from us.

\section{Conclusions}

CN Per 2018 (DN V392 Per) is the fast super-Eddington nova with outburst LC of P-type (smooth decline interrupted by a plateau) according to the  classification [28]. The behaviour of the nova is similar to RN U Sco and extremely fast nova V2672 Oph (suspected RN). All three objects displayed  He/N spectrum classification, large expansion velocities, and triple-peaked emission-line profiles during the decline.
The structure of the expanding shell of CN Per 2018 is similar to the CN V1974 Cygni [3], consisting of an outer fast tenuous low-mass envelope and an inner slow high-mass envelope. The expanding spherical outer envelope and its polar outflows were detected around the maximum. The inner envelope consisting of an equatorial ring and polar blobs was clearly detected 3 days after maximum. The rapid increase of absolute values of radial velocities of polar blobs from  $\sim$ 1750 km/s to $\sim$ 2050 km/s around day 5 after the maximum was caused by a fast bipolar wind from the hot central object after shrinking the pseudophotosphere of the outbursted WD. We used the post maximum spectra of CN Per 2018 and available radio data to estimate the inclination angle of the system i$\sim$ 9$^{\circ}$.

\acknowledgments

We thank to observers R. Leadbeater, E. Bertrand, C. Boussin, J. Edlin, P. Berardi, D. Dejean, J. Montier, O. Garde, M. Verlinden and M. Buchet for the spectra of CN Per 2018 available at the ARAS (Astronomical Ring for Access to Spectroscopy) database. ARAS is an initiative promoting cooperation between professional and amateur astronomers in the field of spectroscopy, coordinated by Francois Teyssier. Our thanks belongs also to R. Kom\v{z}\'ik for 3 spectra obtained at the Skalnat\'e Pleso Observatory. We acknowledge the variable star observations from the AAVSO International Database contributed by observers worldwide and used in this research. This work was supported by the Slovak Research and Development Agency under contract No. APVV-15-0458 and by the Slovak Academy of Sciences grant VEGA No. 2/0008/17.

\end{document}